\documentclass[aps,preprint]{revtex4}%
\usepackage{amsfonts}
\usepackage{amsmath}
\usepackage{amssymb}
\usepackage{graphicx}%
\providecommand{\U}[1]{\protect\rule{.1in}{.1in}}

\begin{document}
\preprint{ }
\title{Information causality and non-locality swapping are equivalent from emergence
of quantum correlations }
\author{Li-Yi Hsu}
\affiliation{Department of Physics, Chung Yuan Christian University, Chungli, 32023,
Taiwan, Republic of China}

\begin{abstract}
Is information causality a new physical principle? To answer this question, we
first analytically derive the criteria of emergence of quantum correlations
from information causality. Then it is shown that, as emergence criteria of
quantum correlations, information causality and uselessness of coupler-based
non-locality swapping can be regarded equivalent. Therefore, incapability of
non-locality swapping using a coupler is as powerful as information causality
in the single-out of quantum physics from generalized non-signalling models.

\end{abstract}
\volumeyear{year}
\volumenumber{number}
\issuenumber{number}
\eid{identifier}
\date[Date text]{date}
\received[Received text]{date}

\revised[Revised text]{date}

\accepted[Accepted text]{date}

\published[Published text]{date}

\startpage{1}
\endpage{7}
\maketitle

\emph{Introduction. \ }Quantum theory and relativity are the two foundations
of modern physics. Quantum theory is non-local, and relativity does not allow
for superluminal signalling. From the perspective of gravity, proving the
coexistence of these two theories is a great challenge for theoretical
physicists. In the non-relativistic case, quantum theory must be a
non-signalling theory, since one can access no information of any distant
party only by local operations. To test the non-local correlations, a lot
of\ Bell-type inequalities are proposed \cite{en}. In general, except
stabilizer-type (or GHZ-Mermin-type) inequalities \cite{GHZ}, a Bell-type
inequality may be violated by an entangled state, but never be violated
maximally by any quantum state. For example, quantum correlations can violate
the celebrated Clauser-Horn-Shimony-Holt (CHSH) inequality \cite{0} up to
2$\sqrt{2}$ rather than 4. As proved by Tsirelson, the amount of non-locality
allowed by quantum mechanics is limited \cite{01}. Furthermore, Popescu and
Rohrlich (PR) pointed out that such a limitation does not result from the
request of relativity \cite{02}. In other words, the CHSH inequality can be
maximally violated even without superluminal signalling, and as a result of
this, an interesting question arises, which is: Is there a more fundamental
principle bounding the non-locality in quantum mechanics?

On the other hand, why not can quantum theory be more non-local? As a
significant property of quantum mechanics, non-locality plays an essential
role in quantum information science. Being regarded as physical resource of
information processing, non-local correlations embedded in quantum
entanglement can be exploited for efficient computation, secure communication,
and some other tasks, such as teleportation \cite{tel} and dense coding
\cite{dc}, which cannot be physically realized without quantum systems. If
quantum correlations were more non-local, quantum information processing would
be more efficient. For example, if the CHSH inequality were maximally
violated, which can be achieved by the PR box (see later), some unconditional
secure computations, such as one-out-of-two oblivious transfer, can be
performed perfectly \cite{15}. In addition, post-quantum correlations take
more advantage in non-local computation than quantum ones \cite{16}. Even in
the noisy PR box, the non-local correlations slightly stronger than quantum
ones would collapse communication complexity, which are strongly believed to
be non-trivial \cite{14}. Therefore, from the perspective of information
science, understanding what bounds the non-local correlations in the quantum
mechanics is also important and interesting.

As previously mentioned, the non-signalling principle cannot be the answer.
Several works focused on the generalized non-signalling models. Wherein, their
probabilistic predictions allow for many common features with quantum
mechanics, such as no-cloning \cite{03,04}, no-broadcasting \cite{04, 041},
and secret correlations \cite{03,06}. In addition, like quantum mechanics, the
non-signalling principle can be exploited for secure key distribution
\cite{07,08}. However, these features do not distinguish quantum correlations
from other non-local correlations \cite{4}. Recently, it was shown that part
of the boundary between quantum and post-quantum correlations partially emerge
from coupler-based non-locality swapping, the analogue of quantum entanglement
swapping \cite{1}. Therein, an imaginary device, called a coupler, is
exploited\ for joint measurements.

Another potential principle is causality. Initially, the relation between
relativistic causality and Tsirelson bounds was investigated \cite{200,2}.
Therein, it is shown that some joint reversible unitary quantum evolutions can
be exploited for entanglement generation and even for signalling.

In this Letter, we focus on another kind of causality: information causality,
which is proposed as a new physical principle by Pawlowski et al. \cite{3}.
Briefly, information causality states that the information gain cannot exceed
the amount of the classical communication, even if non-local correlations
achievable in quantum mechanics can be used as physical resource. On the other
hand, any non-local correlations stronger than quantum ones must violate
information causality. Information causality can therefore be used to specify
quantum theory from other unphysical non-signalling theories.

Is information causality at the very root of quantum theory? The answer is
partially answered by Allock et al.. \cite{4}. Based on the numerical
simulation, it is shown that part of the boundary of quantum correlations
actually emerges from information causality. In this Letter, we study the
relation between information causality and quantum correlations. We
analytically derive the criteria of emergence of quantum correlations from
information causality. Interestingly, information causality, as well as
incapability of non-locality swapping using a coupler \cite{1}, leads to a
quadratic Bell-type inequality. We will show the equivalence between these two
emergence criteria of quantum correlations.

\emph{The bipartite correlation boxes. } \ Before proceeding further, we
briefly review the 8-dimensonal convex no-signalling polytope, which comprises
24 vertices \cite{5,03}. Each vertex corresponds to a bipartite correlation
box (hereafter, just \textquotedblleft box\textquotedblright). Here a box is
defined by a set of two possible inputs for each of spatially separate Alice
and Bob, and a set of two possible outputs for each. Alice's and Bob's inputs
are denoted by $x$ and $y$ respectively, and their outputs by $a$ and $b$. A
joint probability of getting a pair of outputs $a$ and $b$ given a pair of
inputs $x$ and $y$ is $P(a,b|x,y)$, which is definitely positive. In the
following, $a$, $b$, $x$, and $y\in\{0,1\}$. For a non-signalling box,
\[
\sum\nolimits_{b}P(a,b|x,y=0)=\sum\nolimits_{b}P(a,b|x,y=1)=P(a|x),
\]
and%

\[
\sum\nolimits_{a}P(a,b|x=0,y)=\sum\nolimits_{a}P(a,b|x=1,y)=P(b|y).
\]
The probability distribution is unbiased if the marginal probabilities
$P(a|x)=P(b|y)=\frac{1}{2}$, $\forall a$, $b$, $x$, $y$. There are eight
extreme non-local boxes, which have the form:%

\begin{equation}
P_{NL}^{\mu\nu\sigma}(a,b|x,y)=\left\{
\begin{array}
[c]{c}%
\frac{1}{2}\text{ if }a\oplus b=xy\oplus\mu x\oplus\nu y\oplus\sigma\\
\text{0 \ \ \ \ \ \ otherwise, }%
\end{array}
\right.  \label{1}%
\end{equation}
where $\mu$, $\nu$ and $\sigma\in\{0,1\}$ (PR box is the extreme non-local box
with $\mu=\nu=\sigma=0$). The sixteen local deterministic boxes are denoted by
$P_{L}^{\mu\nu\sigma\tau}$, which have the form
\begin{equation}
P_{L}^{\mu\nu\sigma\tau}(a,b|x,y)=%
\genfrac{\{}{.}{0pt}{}{1\text{ if }a=\mu x\oplus\nu\text{ }b=\sigma
y\oplus\tau}{0\text{\ otherwise,}}
\label{0}%
\end{equation}
where $\mu$, $\nu$, $\sigma$ and $\tau\in\{0,1\}$. Given a pair of inputs $x$
and $y$, the corresponding correlator is denoted by%

\[
C_{xy}=\sum\nolimits_{a^{\prime}=b^{\prime}}P(a^{\prime},b^{\prime}%
|x,y)-\sum\nolimits_{a^{\prime}\neq b^{\prime}}P(a^{\prime},b^{\prime}|x,y).
\]
In addition, define $B_{xy}=\left\vert \sum\nolimits_{x^{\prime},y^{\prime}%
}C_{x^{\prime}y^{\prime}}-2C_{xy}\right\vert $ and $\mathbf{B}$ $=\max
\{B_{00}$, $B_{01}$, $B_{10}$, $B_{11}\}$. According to the CHSH scenario, a
mixture of local deterministic boxes must satisfy
\begin{equation}
B_{xy}\leq2,\text{ \ }\forall x,\text{ }y. \label{4}%
\end{equation}
As for an extreme non-local box, we have%

\begin{equation}
\mathbf{B}=4. \label{5}%
\end{equation}
In other words, the CHSH inequalities can be maximally violated by these
non-local boxes. As for the correlations which can be obtained by performing
local measurements on a bipartite quantum system, we have%

\begin{equation}
\mathbf{B\leq}2\sqrt{2} \label{6}%
\end{equation}
In between the non-signalling polytope, the quantum correlations form a body
with a smooth convex curve as its boundary. Given a two-input-two-output
probabilistic distribution, can it be physically realized by quantum systems?
The interesting question was answered independently by Tsirelson, Landau and
Masanes (TLM) \cite{7,8,9}. These proposed criteria on quantumness are
equivalent \cite{10,11}. Here we exploit Landau's criterion, which can be
stated as follows. If a set of correlators $C_{xy}$ is admitted by a quantum
description with unbiased marginals, we have%

\begin{equation}
\mathbf{A}=\left\vert C_{00}C_{10}-C_{01}C_{11}\right\vert \leq\sum
_{j=0,1}\sqrt{(1-C_{0j}^{2})(1-C_{1j}^{2})}. \label{7}%
\end{equation}
It is worthy noting that Navascues et al. recently proposed the criteria on
quantum-obtained correlators $C_{xy}$ with biased marginals \cite{12,13}.

\emph{Information causality}. \ \ Information causality considers the
following communication scenario. Spatially separate Alice and Bob share the
non-local, no-signalling and accessible resources. In addition, Alice is given
$N$ random bits $\overrightarrow{a}=(a_{1}$, $a_{2}$,\ldots, $a_{N})$ and Bob
is given a random variable $b\in\{1,2,\ldots,N\}$. The task for Bob is to
guess the bit $a_{b}$.\ Wherein, Alice is allowed to perform any local
operation on the resource at hand. Then she sends $m$\ classical bits to Bob
via one-way classical communication. As for Bob, he can also perform any local
operation on the accessible resource in his information processing. Finally,
Bob outputs $\mathfrak{b}$ as his answer. To optimize the probability of
successful guessing, the proposed protocol in Ref.\cite{3}\ can be regarded as
the extension of the van Dam's protocol \cite{14}, which is originally
proposed for 1-out-of-2 oblivious transfer \cite{15}.

Information causality is fulfilled if
\begin{equation}
\sum_{K=1}^{N}I(a_{K}:\mathfrak{b}|b=K)\leq m, \label{8}%
\end{equation}
where $I(a_{K}:\mathfrak{b}|b=K)$ is Shannon mutual information between
$a_{K}$ and $\beta$, given $b$ equal to $K$. Equivalently, Eq. (\ref{8}) can
be restated in terms of correlators. That is, information causality is
fulfilled if
\begin{equation}
\mathbf{S}=(C_{00}+C_{10})^{2}+(C_{01}-C_{11})^{2}\leq4. \label{9}%
\end{equation}
Interestingly, Eq. (\ref{9}) is equivalent to the Uffink's inequality, which
is a bipartite quadratic Bell-type inequality \cite{20}. Notably, Eq.
(\ref{9}) cannot be exploited for entanglement testing. It is easy to verify
that $\mathbf{S}$\ ca\bigskip n be maximally violated by PR box. Later it will
be shown that Eq. (\ref{9}) is a weaker form of TLM criteria \cite{11}.

\emph{Emergence of quantum correlations from information causality. \ \ }In
the following, without loss of generality, we assume $C_{00}$, $C_{10}$,
$C_{01}\geq0$ and $C_{11}$ $\leq0$. As a result, $\mathbf{S}$\ can be
optimized in Eq. (\ref{9}) given a set of the absolute values of correlators
$\left\vert C_{xy}\right\vert $. As shown in Fig. 1, in a two-dimensional
Cartesian plane, we define $\overrightarrow{r_{1}}=(C_{00}$, $C_{01})$,
$\overrightarrow{r_{2}}=(-C_{10}$, $C_{11})$, and $\overrightarrow{r_{3}%
}=(C_{11}$, $C_{10})$, which respectively belong to the first, third and
second quadrants. Note that $\overrightarrow{r_{2}}\perp$ $\overrightarrow
{r_{3}}$ and $\left\vert \overrightarrow{r_{2}}\right\vert =\left\vert
\overrightarrow{r_{3}}\right\vert $. The angles between vector pairs
($\overrightarrow{r_{1}}$, $\overrightarrow{r_{2}}$) and hence
($\overrightarrow{r_{1}}$, $\overrightarrow{r_{3}}$) are $\frac{\pi}{2}+\phi$
and $\phi$, respectively. As a result, $\mathbf{S}$ can be regarded as the
area of the square with the side length equal to $\left\vert \overrightarrow
{r_{1}}-\overrightarrow{r_{2}}\right\vert $. On the other hand, $\mathbf{A}$
in Eq. (\ref{7}) is equal to the area of a parallelogram spanned by
$\overrightarrow{r_{1}}$ and $\ \overrightarrow{r_{3}}$ . The relation between
$\mathbf{S}$\ and $\mathbf{A}$ can be revealed by the law of cosine. Namely,%

\begin{align}
\mathbf{S}  &  \mathbf{=}\left\vert \overrightarrow{r_{1}}-\overrightarrow
{r_{2}}\right\vert ^{2}\nonumber\\
=  &  r_{1}^{2}+r_{2}^{2}-2r_{1}r_{2}\cos(\frac{\pi}{2}+\phi)\nonumber\\
&  =r_{1}^{2}+r_{2}^{2}+2r_{1}r_{3}\sin\phi\nonumber\\
&  =r_{1}^{2}+r_{2}^{2}+2A\nonumber\\
&  =\sum_{xy}C_{xy}^{2}+2A, \label{10}%
\end{align}
where $r_{1}^{2}+r_{2}^{2}=\sum_{xy}C_{xy}^{2}$. From Ineq. (\ref{7}), we
have
\begin{equation}
S\leq\sum_{i,j=0,1}C_{ij}^{2}+2\sum_{j=0,1}\sqrt{(1-C_{0j}^{2})(1-C_{1j}^{2})}
\label{11}%
\end{equation}
Now we compare RHS of Ineqs.(\ref{9}) and (\ref{11}). Set $\omega
_{00}=1-C_{00}^{2}$, $\omega_{10}=1-C_{10}^{2}$, $\omega_{01}=1-C_{01}^{2}$,
and $\omega_{11}=1-C_{11}^{2}$, where $\omega_{ij}\geq0$ $\forall i$,
$j\in\{0,1\}$. We have
\begin{align}
&  4-(\sum_{i,j=0,1}C_{ij}^{2}+2\sum_{j=0,1}\sqrt{(1-C_{0j}^{2})(1-C_{1j}%
^{2})})\nonumber\\
&  =\omega_{00}+\omega_{10}+\omega_{01}+\omega_{11}-2\sqrt{\omega_{00}%
\omega_{10}}-2\sqrt{\omega_{01}\omega_{11}}\geq0. \label{12}%
\end{align}
The inequality in (\ref{12}) holds since the arithmetic average is not smaller
than the geometric average. According to Ineq. (\ref{12}), Ineq. (\ref{11}) is
stronger than Ineq. (\ref{9}) and hence Uffink's quadratic inequality is
weaker than TLM criteria \cite{11}.

Furthermore, Ineqs. (\ref{9}) and (\ref{12}) are equivalent when the equality
of the inequality in (\ref{12}) holds. In other words, if
\[
\omega_{00}=\omega_{10}\text{ and }\omega_{01}=\omega_{11},
\]
or equivalently,
\begin{equation}
\overrightarrow{r_{1}}+\overrightarrow{r_{2}}=0\Leftrightarrow C_{00}%
=C_{10}\text{ and }C_{01}=-C_{11}, \label{13}%
\end{equation}
the criteria of the quantumness and information causality in Ineqs. (\ref{9})
and (\ref{11}) coincide. As a result, these two boundaries merge.

Now we consider a mixture of one non-local box and a box $B$ with completely
depolarizing noise as follows \cite{4}%
\begin{equation}
PR_{\lambda,\eta}=\lambda P_{NL}^{\mu\nu\sigma}+\eta B+(1-\lambda
-\eta)\mathbf{1}, \label{14}%
\end{equation}
where $\lambda$, $\eta\geq0$, and $0\leq1-\lambda-\eta\leq1$.

Case (a) : $B$ is a non-local box. That is, $B=P_{NL}^{\mu^{\prime}\nu
^{\prime}\sigma^{\prime}}$. It is easy to verify that $C_{00}=(-1)^{\sigma
}\lambda+(-1)^{\sigma^{\prime}}\eta$, $C_{10}=(-1)^{\mu+\sigma}\lambda
+(-1)^{\mu^{\prime}+\sigma^{\prime}}\eta$, $C_{01}=(-1)^{\nu+\sigma}%
\lambda+(-1)^{\nu^{\prime}+\sigma^{\prime}}\eta$, and $C_{11}$ $=(-1)^{\nu
+\mu+\sigma+1}\lambda+(-1)^{\nu^{\prime}+\mu^{\prime}+\sigma^{\prime}+1}\eta$.
As a result, Eq. (\ref{13}) is satisfied if $\mu=\mu^{\prime}=0$. That is, if
the non-local correlations of $PR_{\lambda,\eta}$ can be obtained by
performing local measurements on a quantum state, the information causality is
automatically fulfilled. Therefore, the boundaries of quantumness and
information causality merge. Obviously, a mixed box $PR=\sum_{\nu,\sigma
}\lambda_{\nu\sigma}P_{NL}^{0\nu\sigma}+(1-\sum_{\nu,\sigma}\lambda_{\nu
\sigma})\mathbf{1}$, where $0\leq\lambda_{\nu\sigma}$ $\forall\nu$, $\sigma$
and $0\leq\sum_{\nu,\sigma}\lambda_{\nu\sigma}\leq1$, can also lead to the
boundary-merging phenomenon.

Case (b) : $B$ is a local box. That is, $B=P_{L}^{\mu^{\prime}\nu^{\prime
}\sigma^{\prime}\tau^{\prime}}$. These four correlators are $C_{00}%
=(-1)^{\sigma}\lambda+(-1)^{\nu^{\prime}+\sigma^{\prime}}\eta$, $C_{10}%
=(-1)^{\mu+\sigma}\lambda+(-1)^{\nu^{\prime}+\sigma^{\prime}+\mu^{\prime}}%
\eta$, $C_{01}=(-1)^{\nu+\sigma}\lambda+(-1)^{\nu^{\prime}+\sigma^{\prime
}+\tau^{\prime}}\eta$, and $C_{11}$ $=(-1)^{\nu+\mu+\sigma+1}\lambda
+(-1)^{\nu^{\prime}+\sigma^{\prime}+\mu^{\prime}+\tau^{\prime}}\eta$. \ The
conditions $C_{00}=C_{10}$and $C_{01}=-C_{11}$ lead to $\mu=\mu^{\prime}=0$
and $\mu=\mu^{\prime}=1$, respectively, which contradict each other.
Therefore, two boundaries never merge in this case.

The reader can refer to Ref. \cite{4} for sketching the slices of the
non-signalling polytope. Wherein, the boundary-merge is demonstrated.

\emph{Discussions. \ \ }In Ref \cite{1}, Skrzypczyk et al. showed the
uselessness of quantum correlations for non-locality swapping. Wherein, the
coupler of two non-local boxes $PR_{\lambda,\eta}=\lambda P_{NL}^{000}+\eta
P_{NL}^{100}+(1-\lambda-\eta)\mathbf{1}$ is studied. The emergence of quantum
correlations from $PR_{\lambda,\eta}$\ uselessness of non-locality leads to a
quadratic Bell-type inequality,
\begin{equation}
(C_{00}+C_{01})^{2}+(C_{10}-C_{11})^{2}\leq4.\label{99}%
\end{equation}
Obviously, Ineq. (\ref{9}) is equal to Ineq.(\ref{99}) under the permutation
of two subscripts. In other words, once Alice and Bob exchange the roles in
the protocol, information causality can therefore be fulfilled by Ineq.
(\ref{99}). On the other hand, as for coupler-based non-locality swapping in
Ref \cite{1}, if the two particles on which the joint measurement is performed
are interchanged with the other two particles, inability of non-locality
swapping is fulfilled by Ineq. (\ref{9}). Information causality and
uselessness of non-locality swapping are therefore equivalent from the
perspective of emergence of quantum correlations. However, the connection
between information causality and non-locality swapping is unclear.

In conclusion, the merging criterion of information causality and quantum
mechanics is analytically derived. These boundaries can be merged with the
unbiased marginals when the quantum correlations can be described by the
depolarized mixture of the $\mu=0$ non-local boxes. In addition, information
causality is equivalent to non-locality swapping from emergence of quantum correlations.

Notably, the proposed protocol in information causality and the scheme for
coupler-based non-locality swapping are asymmetrical. These criteria each
cannot lead to the third quadratic Bell-type inequality, which reads%

\begin{equation}
(C_{10}+C_{01})^{2}+(C_{00}-C_{11})^{2}\leq4.
\end{equation}
Can this inequality be derived from some other principle? It is an open
question. If such principle exists, the employed protocol or scheme should be symmetrical.

The authors acknowledge financial support from the National Science Council of
the Republic of China under Contract No. NSC.96-2112-M-033-007-MY3. This work
is partially supported by the Physics Division of the National Center of
Theoretical Sciences.

\section{Figure Caption:}

Fig. (1) : The geometrical description of $\overrightarrow{r_{1}}$,
$\overrightarrow{r_{2}}$, $\overrightarrow{r_{3}}$, $\mathbf{S}$ , and
$\mathbf{A}$.


\begin{thebibliography}{99}                                                                                               %


\bibitem {en}D. Mermin, Rev. Mod. Phys. \textbf{65}, 803 (1993).

\bibitem {GHZ}M. Ardehali, Phys. Rev. A\textbf{ 46}, 5375 (1992).

\bibitem {0}J. F. Clauser, M.A. Horne, A. Shimony and R. A. Holt, Phys. Rev.
Lett. \textbf{23}, 880-884 (1969).

\bibitem {01}B. S. Tsirelson, Lett. Math. Phys.\textbf{ 4}, 93 (1980).

\bibitem {02}S. Popescu and D. Rohrlich, Found. Phys. \textbf{24}, 379 (1994).

\bibitem {tel}C. H. Bennett, G. Brassard, C. Crepeau, R. Jozsa, A. Peres, and
W. K. Wootters, Phys. Rev. Lett.\textbf{ 70}, 1895 (1993).

\bibitem {dc}C. H. Bennett and S. J. Wiesner, Phys. Rev. Lett. \textbf{69},
2881 (1992).

\bibitem {15}S. Wolf and J. Wullschleger, arXiv:quant-ph/0502030.

\bibitem {16}N. Linden, S. Popescu, A. J. Short, and A. Winter, Phys. Rev.
Lett. \textbf{99}, 180502 (2007).

\bibitem {14}W. van Dam, arXiv:quant-ph/0501159.

\bibitem {03}Ll. Masanes, A. Acin, and N. Gisin, Phys. Rev. A \textbf{73},
012112 (2006).

\bibitem {04}J. Barrett, Phys. Rev. A \textbf{75}, 032304 (2007).

\bibitem {041}H. Barnum, J. Barrett, M. Leifer, and A. Wilce, Phys. Rev. Lett.
\textbf{99}, 240501 (2007).

\bibitem {06}V. Scarani, N. Gisin, and N. Brunner, Phys. Rev. A \textbf{74},
042339 (2006).

\bibitem {07}J. Barrett, L. Hardy, and A. Kent, Phys. Rev. Lett. \textbf{95},
010503 (2005).

\bibitem {08}A. Acin, N. Gisin, and L.Masanes, Phys. Rev. Lett. \textbf{97},
120405 (2006).

\bibitem {4}J. Allcock, N. Brunner, M. Pawlowski, V. Scarani, Phys. Rev. A
\textbf{80}, 040103(R) (2009).

\bibitem {1}P. Skrzypczyk, N. Brunner, and S. Popescu, Phys. Rev. Lett.
\textbf{102}, 110402 (2009).

\bibitem {200}D. Dieks, Phys. Rev. A \textbf{66}, 062104 (2002).

\bibitem {2}H. Buhrman and S. Massar, Phys. Rev. A \textbf{72}, 052103 (2005).

\bibitem {3}M. Paw\l owski, T. Paterek, D. Kaszlikowski, V. Scarani, A.
Winter, and M. \.{Z}ukowski, Nature \textbf{461}, 1101 (2009).

\bibitem {5}J. Barrett , N. Linden, S. Massar, S. Pironio, S. Popescu, and D.
Roberts, Phys. Rev. A \textbf{71}, 022101 (2005).

\bibitem {7}B. Tsirelson, J. Sov. Math. \textbf{36}, 557 (1987).

\bibitem {8}L. Landau, Found. Phys. \textbf{18}, 449 (1988).

\bibitem {9}L. Masanes, quant-ph/0309137.

\bibitem {10}B.S. Tsirelson, Hadronic J. Suppl. \textbf{8}, 329 (1993).

\bibitem {11}A. Cabello, Phys. Rev. A \textbf{72} 012113 (2005).

\bibitem {12}M. Navascues, S. Pironio, and A. Acin, Phys. Rev. Lett.
\textbf{98}, 010401 (2007).

\bibitem {13}M. Navascues, S. Pironio, and A. Acin, New J. Phys. \textbf{10}
073013 (2008).

\bibitem {20}J. Uffink, Phys. Rev. Lett. \textbf{88}, 230406 (2002).
\end{thebibliography}
\end{document}